\documentclass[preprint,11pt]{elsarticle}

\usepackage{amsmath,amsthm}
\usepackage{amssymb,latexsym}
\usepackage{graphicx}
\usepackage{setspace}
\usepackage[mathscr]{eucal}
\usepackage{subfig}

\numberwithin{equation}{section}

\journal{Physica A}

\begin{document}

\begin{frontmatter}


   \author{Haret C. Rosu\fnref{label1}}\ead{hcr@ipicyt.edu.mx, Tel:+524448342000, Fax: +524448342010}
    \author{Jos\'e S. Murgu\'{\i}a\fnref{label2}}
    \author{Andrei Ludu\fnref{label3}}

\title{Scaling analyses based on wavelet transforms for the Talbot effect}

 \address[label1]{IPICYT, Instituto Potosino de Investigacion Cientifica y Tecnologica,\\Apdo Postal 3-74 Tangamanga, 78231 San Luis Potos\'{\i}, S.L.P., Mexico.}

 \address[label2]{Facultad de Ciencias, Universidad Aut\'onoma de San Luis Potos\'i,\\ \'Alvaro Obreg\'on 64, 74800 San Luis
Potos\'{\i}, S.L.P., M\'exico} 
 \address[label3]{Department of Mathematics, Embry-Riddle Aeronautical University,\\ Daytona Beach, FL. 32114-3900, U.S.A.}

\begin{abstract}
\it{
The fractal properties of the transverse Talbot images are analysed
with two well-known scaling methods, the wavelet transform modulus maxima (WTMM) and the wavelet transform multifractal detrended fluctuation analysis (WT-MFDFA). We use the widths of the singularity spectra, $\Delta \alpha=\alpha_{{\rm H}}-\alpha_{\min}$,
as a characteristic feature of these Talbot images. The $\tau$ scaling exponents of the $q$ moments are linear in $q$ within the two methods,
which proves the monofractality of the transverse diffractive paraxial field in the case of these images.}
\end{abstract}


\end{frontmatter}

\newpage

  \section{Introduction} \label{sec-Intro}

The Talbot self-imaging phenomenon discovered in 1836 \cite{ref0}, is a near-field diffraction effect with many potentially important technological applications \cite{P89}. It has been explained as a constructive interference by Lord Rayleigh in 1881 \cite{R81}, who was able to show that the Talbot images are formed behind a coherently illuminated diffraction grating at even multiples of the Talbot distance $z_T=a^2/\lambda$, where $a$ is the period of the grating and $\lambda$ the wavelength of the light.
For a binary transmission (Ronchi) grating and in the paraxial approximation, Winthrop and Worthington have shown that superpositions of $r$ equally-spaced clear images of the grating are formed at any fractional distance  $z=p/r$, where $p$ is coprime with $r$ \cite{refa}.
In 1996, in the framework of Helmholtz equation approach to the physical optics of a $\delta$-comb grating, the results of Winthrop and Worthington have been confirmed and substantially extended by Berry and Klein \cite{ref1}. They showed that in rational Talbot planes, i.e., planes situated at any rational distances in units of the Talbot distance, the paraxial diffraction wavefield has an interesting arithmetic structure related to Gauss sums in number
theory. Moreover, they showed that at irrational distances a fractal structure of the diffraction field can be proved with sufficient
experimental accuracy.


A few years ago, we already applied wavelet transforms (WTs) to the Talbot effect \cite{rosuT1}, but the arguments were based on
the logarithmic plot of the variance of the wavelet coefficients as a function of the wavelet level $m$. It was found that the latter plot is linear for the standard Talbot effect confirming the results of Berry and Klein. However, it is known that monofractals can be also characterized through multifractal spectra 
of narrow support \cite{Drozdz09} and more generally the multifractal scaling analyses can be used to disentangle multifractal behaviour from monofractal one \cite{K01}. This motivates us to apply here two of the most efficient multifractal methods, the WTMM and the WT-MFDFA, to the scalar paraxial diffraction field. The main difficulty when applying these methods is due to the fact that the side of negative momenta of the singularity spectra is usually sensitive to measurement noise and window quantization which generate a rather strong bias of the estimators based on wavelets \cite{VF09}.
Thus, we employ the supports of the singularity spectra from $\alpha_{\min}$ up to the limiting singularity, $\alpha_{\rm H}$, which defines the Hurst exponent, to characterize the Fibbonaci convergent Talbot images and we also show their sensitivity to the positions at which the images are formed.


The paper is organized in the following way. 
In Section~\ref{sec-T2}, we briefly describe the fractal Talbot effect as introduced by Berry and Klein. Section~\ref{sec-T3} contains
a brief review of the two employed wavelet methods followed in Section~\ref{sec-T4}
by their application to the intensity signal of the paraxial diffraction field for several distances belonging to the $\bar{\zeta}_{{\rm G}}$ Fibonacci sequence in the Talbot distance units. Furthermore, in the WT-MFDFA method, we calculate the Hurst exponent from the logarithm of the fluctuation function $F_q$ in equation~(\ref{eq-Fqs}) below and use it to obtain the singularity spectra in an independent different way. Some conclusions end up the paper.

  \section{The Fractal Analysis of Berry and Klein} \label{sec-T2}
%

The following expression for the paraxial diffracted wave field behind a Ronchi grating
\begin{equation}\label{bk24}
\Psi_{{\rm p}}(\xi , \zeta)=\frac{1}{2}+\frac{2}{\pi}\sum_{k=0}^{\infty}\frac{(-1)^k e^{-i\pi(2k+1)^2\zeta}}{2k+1}\cos[2\pi
(2k+1)\xi] 
\end{equation}
has been obtained by Berry and Klein who also proved that its square, i.e., the paraxial light intensity, is an anisotropic fractal in this case. Their findings were based on the general result that the plot of a function with power spectrum $|g_n|^2$ proportional to $n^{-\gamma}$ is a fractal curve with fractal dimension $D=(5-\gamma)/2$ and $\gamma$ is the spectral parameter of the wave signal. Indeed, as a consequence of this argument, they showed that in the transverse planes perpendicular to the incident light located at irrational values of $\zeta = z/z_T$ the intensity of the Talbot images
is a fractal signal whose graph has dimension $\frac{6}{4}$, while in the longitudinal planes parallel to the
incident light and {\em almost} all oblique planes, the intensity is a fractal whose graph has dimension $\frac{7}{4}$. Finally, in certain {\em special} diagonal planes, the fractal dimension has a minimum value of $\frac{5}{4}$.
In this work, we will focus on the transverse Talbot effect since it is the most relevant one being directly observed on any screen perpendicular to the direction of propagation. However, the present wavelet analyses can be easily extended to the other Talbot fractals introduced by Berry and Klein and to the probability density landscapes in quantum mechanics \cite{b275}, also known as {\em quantum carpets}, and more generally to any kind of revival patterns in wave physics \cite{kaplan-2000}.
Since in the transverse case $\gamma=2$, which is related to the Hurst exponent ${\rm H}$ by $\gamma=2{\rm H}+1$, we see that we are in the case of uncorrelated fractal with ${\rm H}=0.5$, a value that we confirmed with WTMM in \cite{rosuT1} and reconfirm with the WT-MFDFA method in Fig.~1, where the scaling exponent $\alpha$ is related with the spectral exponent $\gamma$ by $\alpha = (1+\gamma)/2$.

\medskip

To obtain the fractal Talbot images, Berry and Klein considered the irrational $\zeta _{\rm irr}$ as the limit $m \rightarrow \infty$ of
a sequence of rationals $p_m/r_m$. In particular, they employed the successive truncations (also known as convergents) of the continued fraction expression for $\zeta _{\rm irr}$, namely
\begin{equation}\label{bk25}
 \zeta _{\rm irr} =a_0+\frac{1}{a_1+\big[\frac{1}{(a_2+\cdot \cdot \cdot)}\big]}~,
\end{equation}
where the $a_m$ are positive integers. These sequences give best approximations, in the sense that $p_m/q_m$ is closer to $\zeta
_{\rm irr}$ than any other fraction with denominator $r\leq r_m$. As a matter of fact, they considered the inverse golden mean $\zeta _{\rm
G}^{-1}=(5^{1/2}-1)/2$, for which $a_0=0$ and all the other $a_m$ are unity, and the $p_m$ and $r_m$ are Fibonacci numbers. Because of the symmetries of the paraxial field, the Talbot images are the same at $\zeta_{{\rm G}}^{-1}$ and $\bar{\zeta}_{{\rm G}}\equiv 1-\zeta_{{\rm G}}^{-1}$. The latter has the following numerical value
$$
\bar{\zeta}_{{\rm G}}=\frac{3-5^{1/2}}{2}=0.381966...
$$
and the set of convergents:
\begin{equation}\label{bk26}
\Big\{
\frac{1}{2}~,\,\frac{1}{3}~,\,\frac{2}{5}~,\,\frac{3}{8}~,\frac{5}{13}~,\,\frac{8}{21}~,\, \frac{13}{34}~,\,\frac{21}{55}~,
\,\frac{34}{89}~,\,\frac{55}{144}~,\,\frac{89}{233}~,\,...\Big\}~.
\end{equation}
We denote the fractions in this set as $Q_1=1/2$, $Q_2=1/3$, and so forth. The odd ones are to the right of $Q_{\infty}=0.381966$ with the biggest at 0.5, whereas the even ones are to the left with the smallest at 0.(3). Thus they cover an asymmetric range of 0.1666 around their limit. The two wavelet analyses will be applied to the transverse diffracted field at these distances but other fractional distances outside this interval could be used as well as briefly commented at the end of the paper.

\section{WT, WTMM, and WT-MFDFA} \label{sec-T3}

{\em WT} -- Among the numerous integral transforms for signal processing the {\em wavelet transforms} include the scale as one of their arguments and therefore provide in a natural way detailed information on the fractal-type structures \cite{Mallat}.
These transforms make usage of integration kernels called wavelets and for a given signal $y(x)$ they are expressed as follows
\begin{equation}\label{eq-CWT}
W_y(a, b) = \frac{1}{\sqrt{a}} \int_{-\infty}^{\infty} y(x)
\bar{\psi} \left(\frac{x - b}{a} \right) dx,
\end{equation}
where $\psi$ is the analyzing wavelet,
$b \in \mathbf{R}$ is a translation parameter, whereas $a \in \mathbf{R}^{+} ~ (a \neq 0)$ is a
dilation or scale parameter, and the bar symbol denotes complex conjugation.

To be able to analyze appropriately the singular behavior, the wavelets $\psi(x)$ should have enough vanishing moments~\cite{Bacry93, Mallat}.
A wavelet is said to have $n$ vanishing moments if and only if it satisfies
$\int_{-\infty}^{\infty} x^k \psi(x) dx = 0, \quad {\rm for} ~ k = 0, 1, \ldots , n - 1,\,
\quad {\rm and} \quad \int_{-\infty}^{\infty} x^k \psi(x) dx \neq 0~ \quad {\rm for} ~ k \geq n$.
This means that a wavelet with $n$ vanishing moments is orthogonal to all polynomials of degrees up to $n-1$. Thus, the wavelet
transform of $y(x)$ performed by means of a wavelet $\psi(x)$ with $n$ vanishing moments is a ``smoothed version'' of the $n$th derivative of $y(x)$ on various scales.


In the presence of isolated singularities at particular points $x_{0i}$, $i=1,2,...,n$, the scaling behavior of the wavelet transforms in the small scale limit $a \to 0^+$ is described by power laws with exponents $\alpha(x_{0i})$ known as H\"older exponents,

\begin{equation} \label{eq-singular}
 W_y(a, x_{0i}) \sim a^{\alpha(x_{0i}) + 1/2}~.
\end{equation}
{\em WTMM} -- To characterize the singular behavior of signals, or in general any functions, and to reveal the hierarchical structure of the singularities, it is sufficient to consider the values and positions of WTMM \cite{mallat-huang} defined as points $(a_0, b_0)$ on the scale-position plane, $(a,b)$, where $|W_y(a_0, b)|$ is locally maximum for $b$ in the neighborhood of $b_0$. These maxima are located along curves in the plane $(a,b)$ known as maxima lines.
The wavelet multifractal formalism may characterize fractal objects which cannot be completely described using a single fractal dimension.
Bacry {\em et al} \cite{Bacry93} defined an ``optimal'' partition function ${\mathcal Z}_q(x, a)$ in terms of the WTMM.
 They considered the set of modulus maxima at a scale $a$ as a covering of the singular
 support of $x$ with wavelets of scale $a$. The partition function ${\mathcal Z}_q(x, a)$ measures the sum of all wavelet modulus maxima at a given power $q$ as follows
\begin{equation} \label{Mallat:FP}
 {\mathcal Z}_q(x, a) = \sum_s |W_y(a, b_s(a)) |^q,
\end{equation}
where $\{ b_s(a) \}_{s \in \mathbf{Z}}$ is the set of positions of all local maxima of $|W_y(a, b)|$ at a fixed scale $a$.
It can be inferred from (\ref{Mallat:FP}) that for $q>0$ the most pronounced modulus maxima will prevail, whereas for $q<0$ the lower ones will survive. For each $q \in \mathbf{R}$, the partition function 
 is related to its scaling exponent $\tau(q)$ through ${\mathcal Z}_q(x, a) \sim a^{\tau(q)}$.
 A linear behavior of $\tau(q)$ indicates monofractality whereas nonlinear behavior suggests that a signal is a
 multifractal. A fundamental result in the wavelet fractal formalism states that the singularity (H\"older) spectrum $f(\alpha)$ of
 the signal $y(x)$ is the Legendre transform of $\tau(q)$, i.e.,
\begin{equation} \label{Leg:D-tau}
\alpha(q) = \frac{d\tau(q)}{dq} \qquad {\rm and} \qquad  f(\alpha)  = q\alpha - \tau(q).
\end{equation}
The H\"older spectrum of dimensions, $f(\alpha)$, is a non-negative convex function that is supported on
the closed interval $[\alpha_{{\rm min}}, \alpha_{{\rm max}}]$, which is interpreted as the Hausdorff fractal dimension of
 the subset of data characterized by the H\"older exponent $\alpha$ \cite{Salo1}.
The most ``frequent'' singularity, which corresponds to the maximum  of $f(\alpha)$, occurs for the value of $\alpha(q=0)$, whereas the boundary
values of the support, $\alpha_{{\rm min}}$ for $q>0$ and $\alpha_{{\rm max}}$ 
correspond to the strongest and weakest singularity, respectively.

 The analyzing wavelets which are used most frequently are the successive derivatives of the Gaussian function
  \begin{equation}\label{eq-Wavelets}
      \psi^{(n)}(x) := \frac{d^n}{dx^n}\left( \exp(-x^2 / 2)\right), \qquad n \in \mathbf{Z}^+,
  \end{equation}
 because they are well localized both in space and frequency, and they remove the trends of the signal that can be approximated by polynomials
 up to $(n- 1)$th order. 
 In particular, our analyses were carried out with the Mexican hat wavelet $\psi^{(2)}(x)$ because we have found that the usage of higher derivatives does not change the results in a significant way.

\medskip

\noindent {\em WT-MFDFA} -- This method that we also used in previous papers \cite{pps1,pps2} is based on the calculation of the so-called $q$th order fluctuation function defined as
          \begin{equation}\label{eq-Fqs}
            F_q(s;m) = \left\{ \frac{1}{2M_s} \sum_{\nu=1}^{2M_s} |F^2(\nu,s;m)|^{q/2} \right\}^{1/q}~,
          \end{equation}
where $q \in \mathbf{Z}$ with $q \neq 0$ and $F^2(\nu,s;m)$ is the standard deviation of the cumulative signal in window $\nu$ of length $s$ at level $m$, for more details see \cite{pps1,pps2}. The divergent behaviour for $q\to 0$ can be avoided by using
the logarithmic averaging $F_0(s;m) = \exp\left\{\frac{1}{2M_s} \sum_{\nu=1}^{2M_s} \ln |F^2(\nu,s;m)| 
\right\}$. 
 We have found that a better matching of the results given by the WT-MFDFA method with those of other
methods is provided by the Db-4 wavelets with four filter coefficients.

If the fluctuation function $F_q(s;m)$ displays a power law scaling
       \begin{equation}\label{eq-FqsPLaw}
         F_q(s;m) \sim s^{h(q)},
       \end{equation}
then the analyzed data have a fractal scaling behaviour. The exponent $h(q)$ is the generalized Hurst exponent since it can depend on $q$,
while the original Hurst exponent is ${\rm H}=h(2)$.
Thus, $h(q)$ can be determined from the double logarithmic plot of the fluctuation function.
If $h(q)$ is constant for all $q$ then the time series is monofractal, otherwise it has a MF behavior. One can also calculate various other scaling exponents, such as $\tau(q)$ and $f(\alpha)$. 


\section{Application to the transverse Talbot effect} \label{sec-T4}
We present now our results on the transverse Talbot effect concerned with the two WT techniques together with the conclusions that come out of these analyses.

The results of the WTMM and WT-MFDFA methods presented in Figure~\ref{fig-mf2} correspond to the distance 
$Q_{11}=89/233$ that has been chosen for illustrative purposes from a set of similar plots that we obtained for over 20 Fibonacci fractional distances. In the figure, we present the light intensity, i.e., the modulus square $|\Psi_{{\rm p}}|^2$ of the Ronchi signal given by the sum in Eq.~\ref{bk24} truncated at a sufficiently large order, for $\xi$ in the unit cell and fixed $\zeta=Q_{11}$, 
 its WTMM graph in semilogarithmic wavelet-parameter plane, and the resulting positive sides of the singularity spectrum.

For comparison purposes, we present the positive sides of the singularity spectra and the resulting $\tau$ exponents obtained through the two methods in Figs.~\ref{fig-mf-3sp} and \ref{fig-mf-3tau}. From these figures one can conclude that the WT-MFDFA method is much more stable numerically.


\medskip

We characterize the fractal behavior of the near-field diffraction with the width $\Delta \alpha =  \alpha_{\rm{H}} - \alpha_{\rm{min}}$ of the positive $q$ values of the parabolic singularity spectrum $f(\alpha)$. 
Thus, in Fig.~\ref{fig-mf5}, we display the dependence of the widths $\Delta \alpha$ on the convergent in the Fibonacci sequence of scaled distances $\zeta$. We notice that $\Delta \alpha$ goes quite rapidly to the value of $\approx$ 0.2551 that we take as the limiting width corresponding to the irrational scaled distance of the inverse golden mean value. We find again that the convergence is better, in the sense of less errors, in the case of WT-MFDFA as compared with WTMM.

The mean values of the widths are $\overline{\Delta \alpha} = 0.2447\,(0.2568)$ when the WTMM (WT-MFDFA) methods are employed.
We also notice that the mean values of the most ``frequent'' singularity for all analysed sequences occur at $\bar{\alpha}_{{\rm H}} = 0.6040\,(0.5058)$.
Once again, these values favor the WT-MFDFA with respect to the WTMM.

Examining the possibility noticed in the literature \cite{GP1,GP2,GM} that finite size effects could introduce spurious multifractal effects, we have found a small decreasing trend of $\overline{\Delta \alpha}$ with increasing length of the signal in the case of WT-MFDFA. In general, the decrements are at the level of -0.01 for our data for lengths of the signal from $L=2^9$ to $L=2^{15}$. On the other hand, in the case of the WTMM method the effect is opposite and more defined and could be due to the correlations introduced by the usage of the lines of modulus maxima. It appears that the WT-MFDFA shows more robustness to residual multifractal effects than WTMM.

\medskip

Finally, we want to draw attention to an important issue that we found by using the multifractal analyses, namely that in principle they can be used for precise monitoring of the distances in optical setups. Let us take the fraction $\frac{113}{355}=0.3183$ which although made of coprime numbers does not belong to the Fibonacci sequence considered here. The results of the WTMM and WT-MFDFA methods for this fraction are shown in Fig.~\ref{fig-mf6}. Although $\alpha_{{\rm H}}$ is very close to 0.5, one can notice a clear difference in the the pattern of the transformed signal with respect to the Fibonacci convergents. The investigation of this issue is beyond the goals of this work.

In conclusion, we have shown that scaling methods based on wavelets are useful techniques in the study of near-field optical diffraction in the particular case of one of its most relevant effects. We have also found in our application that the conclusions of the comparative study of O\'swie\c cimka {\em et al} \cite{OKD} between MFDFA and WTMM methods are also valid when WT-MFDFA usage is compared to that of WTMM.

%
 \begin{figure} [x]
  \centering
  \includegraphics[width= 10.15 cm, height=14 cm] {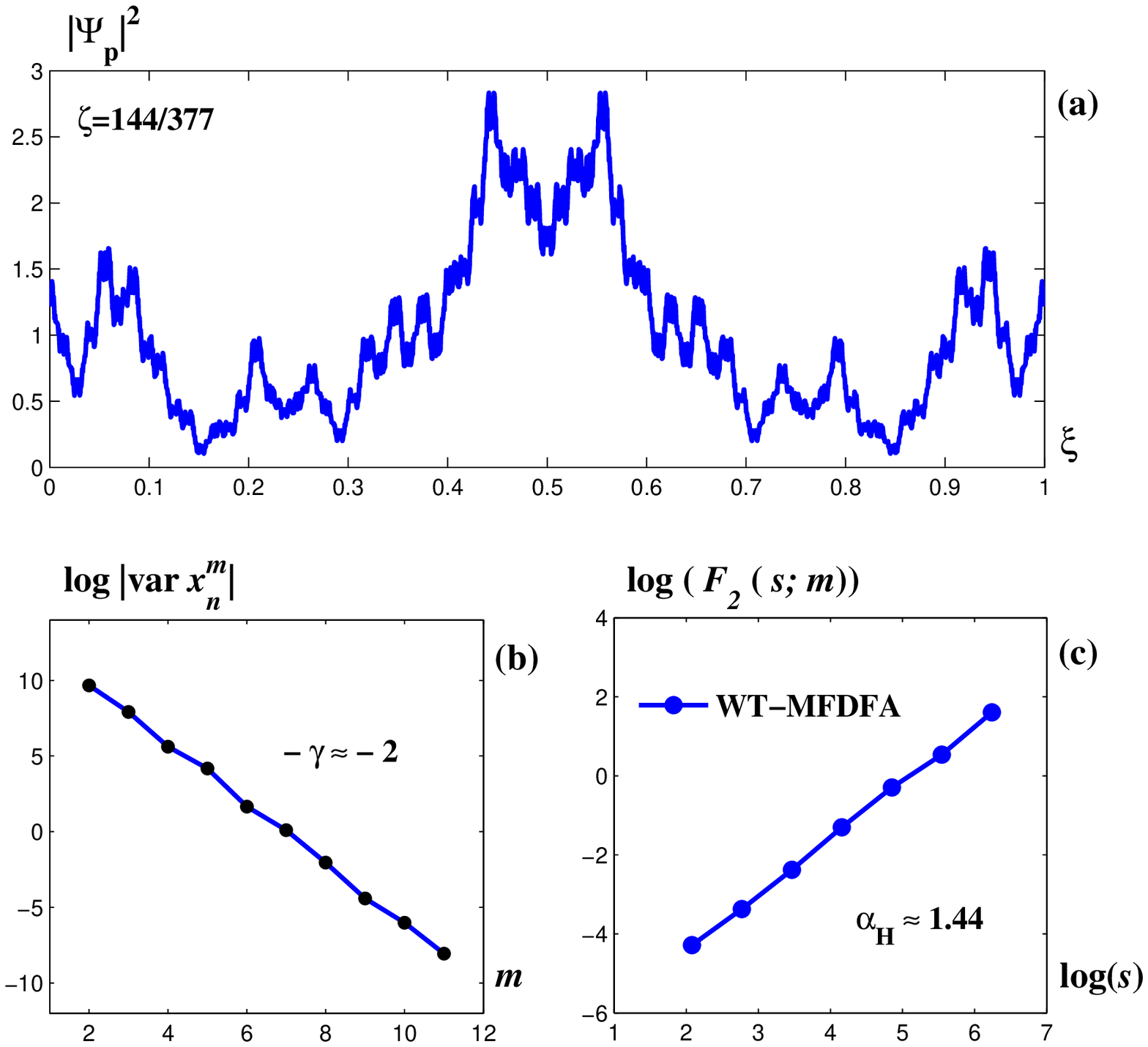}
  \caption{(Color online) The transverse Talbot image (a), the plot of the logarithm of the wavelet coefficients versus wavelet levels (b), and the fluctuation function $F_2$ of the WT-MFDFA method (c), for $Q_{12}$. The Hurst exponent provided by $F_2$ is 1.44, which is close to 1.5 that comes from $\gamma=2$.}
  \label{fig-mf1}
 \end{figure}
%


 \begin{figure}[x] 
  \centering
  \includegraphics[width= 10.15 cm, height=14 cm]{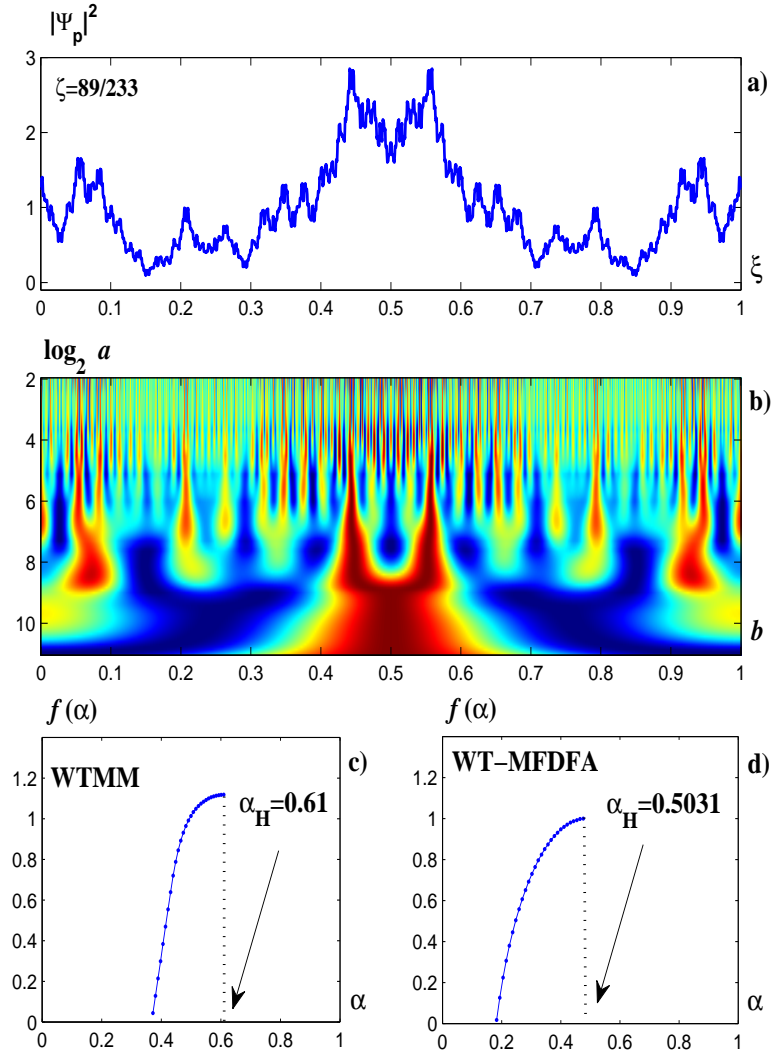}
  \caption{(Color online) Scaling analyses for the Talbot image signal at $Q_{11}$: a) The transverse image intensity, b) its WTMM graph, and  c) the positive $q$ sides of the singularity spectrum obtained with the WTMM and WT-MFDFA methods.}
  \label{fig-mf2}
 \end{figure}


 \begin{figure} [x]
  \centering
  \includegraphics[width= 10.64 cm, height=9.08 cm] {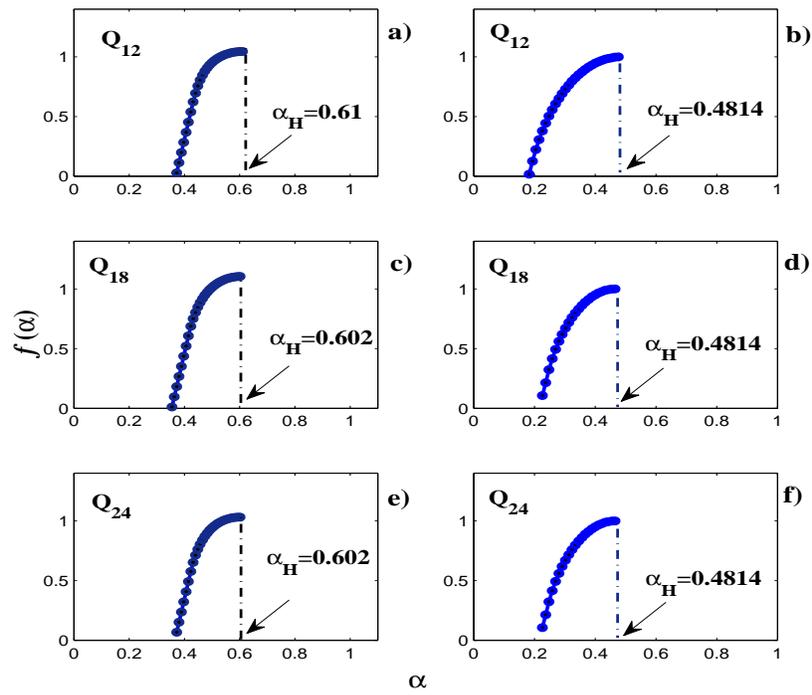}
  \caption{(Color online) The positive $q$ sides of the singularity spectra for three $Q$ distances with (a) the WTMM method, and (b) the WT-MFDFA method.}
  \label{fig-mf-3sp}
 \end{figure}


 \begin{figure} [x]
  \centering
  \includegraphics[width= 10.64 cm, height=9.08 cm]{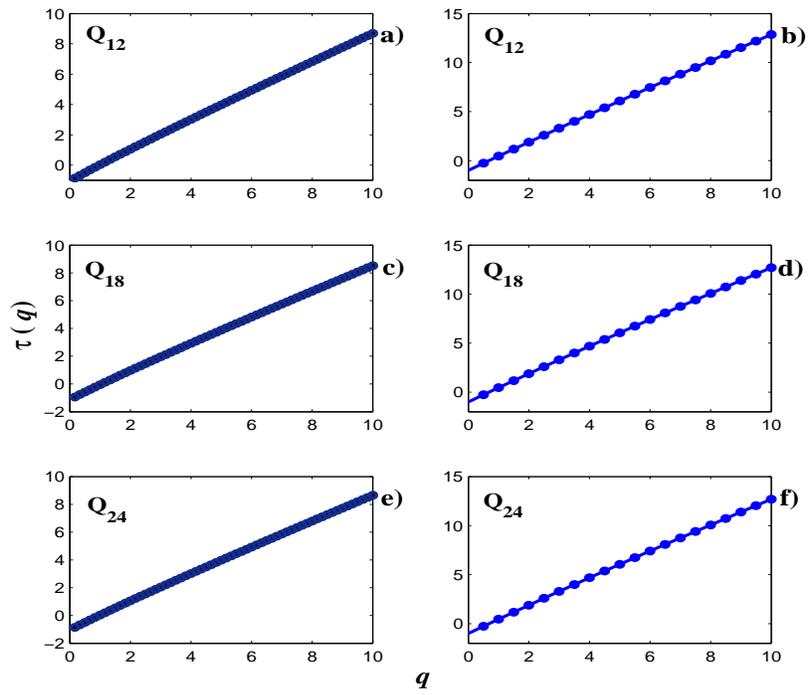}
  \caption{(Color online) The $\tau$ plots for the same three $Q$ distances with the WTMM method (left column), and the WT-MFDFA method (right column).}
  \label{fig-mf-3tau}
 \end{figure}

 \begin{figure}[x]
  \centering
  \includegraphics[width= 7.08 cm, height=13.07 cm]{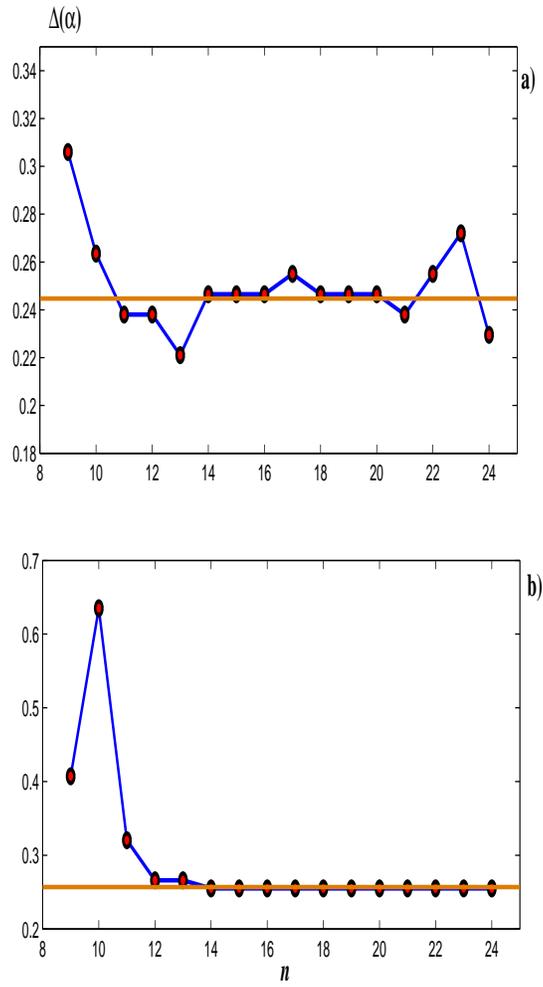}
  \caption{(Color online) $\Delta \alpha=\alpha_{{\rm H}}-\alpha_{min}$ versus the position of the Fibonacci fraction in the Fibonacci sequence obtained through (a) the WTMM method and (b) the WT-MFDFA.}
  \label{fig-mf5}
 \end{figure}

 \begin{figure}[x]
  \centering
  \includegraphics[width= 7.08 cm, height=13.07 cm]{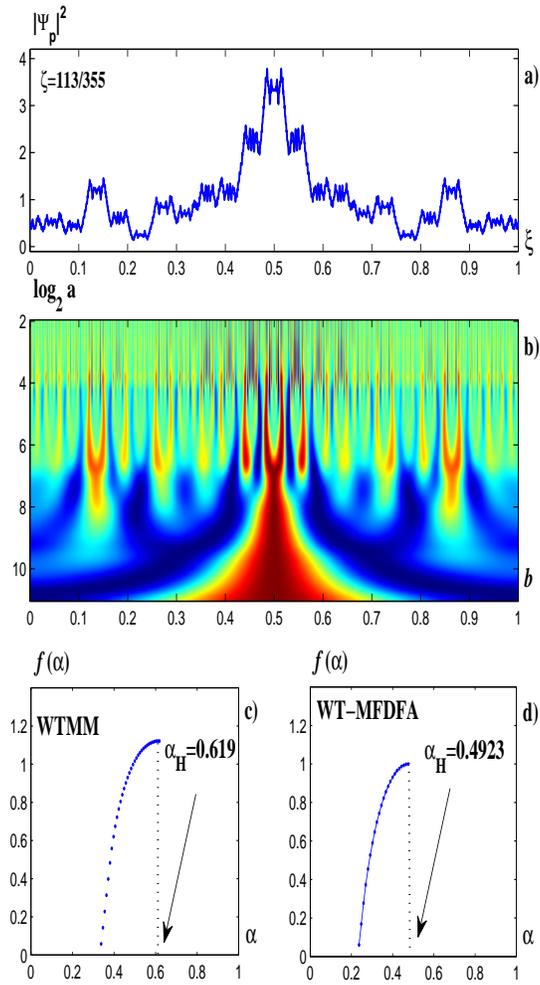}
  \caption{(Color online) WTMM and WT-MFDFA analyses for the fractional distance 113/355 which does not belong to the Fibonacci sequence for $\bar{\zeta}_{{\rm G}}$.}
  \label{fig-mf6}
 \end{figure}

%

\end{document}